\documentclass[a4paper,fleqn]{cas-dc}

\usepackage[authoryear]{natbib}
\usepackage{rotating}
\usepackage{siunitx}
\usepackage[switch]{lineno}
\usepackage{stfloats}

\def\tsc#1{\csdef{#1}{\textsc{\lowercase{#1}}\xspace}}
\tsc{WGM}
\tsc{QE}
\tsc{EP}
\tsc{PMS}
\tsc{BEC}
\tsc{DE}

\begin{document}
\let\WriteBookmarks\relax
\def\floatpagepagefraction{1}
\def\textpagefraction{.001}
\shorttitle{Determining elemental composition in laboratory meteorite ablation spectra through radiative transfer modeling}
\shortauthors{A. Pisarčíková et~al.}

\title [mode = title]{Determining elemental composition in laboratory meteorite ablation spectra through radiative transfer modeling}                      



\author[1]{Adriana Pisarčíková}[type=editor,
                        auid=000,bioid=1,
                        orcid=0000-0002-4084-1248]
\cormark[1]
\ead{adriana.pisarcikova@asu.cas.cz}

\credit{Conceptualization, Investigation, Data curation, Formal analysis, Writing - Original draft preparation}

\affiliation[1]{organization={Astronomical Institute of the Czech Academy of Sciences},
                addressline={Fričova 298}, 
                city={Ondřejov},
                postcode={25165}, 
                country={Czech Republic}}

\author[1]{Jiří Borovička}

\credit{Methodology, Software, Writing – review \& editing}

\author[2]{Pavol Matlovič}[%
   ]

\credit{Conceptualization, Investigation, Data curation}

\affiliation[2]{organization={Faculty of Mathematics, Physics and Informatics, Comenius University Bratislava},
                addressline={Mlynská dolina}, 
                postcode={84248}, 
                postcodesep={}, 
                city={Bratislava},
                country={Slovakia}}

\cortext[cor1]{Corresponding author}


\begin{abstract}
Laboratory simulations of meteor ablation provide a critical quantitative link between the chemical composition of meteoroids and their observed spectral features. In this work, we analyzed high-resolution Echelle spectra (wavelength range 380–780 nm) of 22 diverse meteorites from the dataset presented in our previous work \citep{2024A&A...689A.323M}, representing the largest collection of laboratory meteor analogs to date. Using a radiative transfer model assuming local thermodynamic equilibrium (LTE) and accounting for self-absorption in optically thick plasma, we derived plasma parameters and elemental abundances for both major (Fe, Mg, Cr, Mn, Si, Na, Ni, Li, and K) and minor (Co, Cu, and V) species. Comparison with known bulk meteorite compositions allowed us to validate the modeling approach and assess chemical biases resulting from laboratory-induced ablation. Our analysis suggested plasma temperatures between 5220 and 5810 K and revealed systematic discrepancies in the elemental abundances compared to the original chemical composition. Specifically, we observed a significant enhancement of volatile species (Na, K) relative to Fe, accompanied by a depletion of the moderately volatile element Mg, while refractory elements (Al, Ca, Ti) remained undetected in the plasma radiation. These trends are consistent with the equilibrium vaporization model and demonstrate that under the simulated entry conditions ($\sim$12\,km s\textsuperscript{-1} at $\sim$80\,km altitude), the ablation process is dominated by incomplete and fractional vaporization. We conclude that while laboratory spectra of plasma from ablated meteorites do not fully reflect the original bulk composition, radiative transfer modeling effectively characterizes the state of the radiating plasma, offering a more robust approach for interpreting compositional properties from meteor observations.

\end{abstract}


\begin{highlights}
\item Radiative transfer analysis of ablation spectra from 22 diverse meteorite types
\item Laboratory spectra are consistent with optically thick plasma radiation
\item Calculated elemental abundances reveal incomplete vaporization
\end{highlights}


\maketitle

\section{Introduction}

Meteoroids represent a significant component of interplanetary material, offering critical insights into the origin, dynamical evolution, and compositional history of Solar System bodies. Analysis of their chemical composition is essential for mapping the distribution of primitive and differentiated materials, as well as constraining the physical and chemical conditions in the early protoplanetary disk. Meteor emission spectra, recorded during atmospheric entry of meteoroids from various regions of the Solar System, currently constitute the primary source of information on meteoroid elemental composition through ablation-induced release of atomic and molecular species into radiating plasma \citep{1998SSRv...84..327C}. However, the interpretation of meteor spectra is indeed complex, since the observed line intensities are affected mainly by plasma thermodynamics, self-absorption effects, the volatility and refractory nature of individual elements, and the temporal evolution of the ablating material \citep{1993A&A...279..627B, 2004EM&P...95..413S}. Consequently, deriving accurate compositional information from meteor spectra requires methods capable of linking the observed radiative properties to the real chemical composition of the meteoroid.

Analysis of meteor emission spectra commonly uses line intensity ratios as markers of elemental abundances, yielding valuable qualitative information about spectral differences among meteors of different sizes (e.g. \citet{ 2005Icar..174...15B, 2019A&A...621A..68V, 2019A&A...629A..71M, 2020P&SS..19405040A, 2022MNRAS.513.3982M}). However, systematic discrepancies arise when correlating observed spectral data directly with meteorite bulk composition, highlighting the critical gap in linking real meteoroid material to its emission spectrum. To address these challenges, recent years have seen a growing trend toward laboratory simulations of meteor ablation using meteorites to generate artificial meteors. These experiments benefit from a known initial composition of the samples, controlled physical parameters, and reproducible measurements, enabling deeper interpretation of ablation spectra. Various approaches have been explored: \citet{2019A&A...630A.127F} generated plasma analogous to meteor ablation using high-power terawatt-class laser-induced breakdown spectroscopy; \citet{2021ExA....51..425K} employed dielectric breakdown induced by high-power lasers to study ablation processes; a series of experiments at the IRS (Institute of Space Systems) plasma wind tunnel facility, starting with the first meteorite ablation tests by \citet{2017ApJ...837..112L}, followed by further analyses of several meteorite types by \citet{2018A&A...613A..54D}, and leading to the extensive collection of tested meteorite samples presented in \citet{2024A&A...689A.323M}, were all conducted under the same flow conditions; and finally, \citet{2019ApJ...876..120H} investigated two meteorite analogs in a high-enthalpy facility using the VKI Plasmatron at the Von Karman Institute for Fluid Dynamics. In \citet{2024A&A...689A.323M}, we have presented the largest set of high-resolution laboratory spectra of various meteorite types to date and demonstrated that line-intensity ratio variations from artificial meteors can significantly improve meteoroid classification. While that work focused on qualitative interpretation of spectral features suitable for distinguishing compositional types from line intensity measurements in meteors, it did not include comprehensive radiative transfer modeling required for direct comparison with the bulk elemental ratios of the tested meteorites.

In this work, we overcome this gap by modeling laboratory meteorite ablation spectra to derive plasma conditions and elemental abundances, thereby providing a quantitative link between material composition and its observed spectral signatures. We fitted high-resolution Echelle spectra of 22 diverse meteorites presented in our work in \citet{2024A&A...689A.323M} using a radiative transfer model assuming local thermodynamic equilibrium (LTE) and accounting for self-absorption in optically thick plasma \citep{1993A&A...279..627B}. Such an approach has so far been applied only to a limited number of individual fireball spectra \citep{1993A&A...279..627B, 2003M&PS...38.1283T, 2007AdSpR..39..491J, 2018A&A...610A..73F, 2022MNRAS.514.5266K}, but it has not yet been validated on laboratory spectra of meteorites with known bulk composition. By performing such validation for a diverse set of meteorite types, we are able to quantitatively assess the reliability of the model, identify systematic effects associated with laboratory ablation, and evaluate the extent to which derived elemental abundances reflect the original material composition.

First, in Section \ref{Instrumentation}, we describe the instrumentation, experimental setup, and the methodology used for spectral data modeling and abundance determination. Section \ref{Results} presents the results, including the derived plasma parameters and relative elemental abundances from radiative transfer modeling of spectra from laboratory tested meteorites. This section also provides a comparison with the sample bulk composition and discusses the results in the context of an equilibrium vaporization model. Finally, the conclusions derived from the obtained results are summarized in Section \ref{Summary}.

\section{Instrumentation, data and methodology}
\label{Instrumentation}

\subsection{Experimental setup and data acquisition}

The spectral data analyzed in this study originate from laboratory ablation experiments conducted in the plasma wind tunnel facility PWK1 of IRS at the University of Stuttgart, Germany. These experiments, performed within the MetSpec project and fully described in \citet{2024Icar..40715791T, 2024A&A...689A.323M, 2024Icar..40715817L, 2023Icar..40415682P}, and references therein, simulate the atmospheric entry of slow asteroidal meteoroids under controlled conditions. Meteorite samples were exposed to a high-enthalpy plasma flow with a local mass-specific enthalpy of 70 MJ kg\textsuperscript{-1} and a stagnation pressure of $\sim$24\,hPa, corresponding to the atmospheric entry of a $\sim$4\,cm body at $\sim$80\,km altitude with an assumed entry velocity of $\sim$12\,km s\textsuperscript{-1} \citep{2024Icar..40715817L}. In total, 22 meteorite samples of various compositional classes were tested, forming the largest dataset of laboratory meteor analogs available to date.

The emission spectra were recorded with the high-resolution Echelle spectrometer LTB ARYELLE 150 operated by the High Enthalpy Flow Diagnostics Group (HEFDiG) at IRS \citep{2024Icar..40715817L}. This fiber-fed instrument covers the 250–880 nm wavelength range with a spectral dispersion that varies from 43 pm px\textsuperscript{-1} to 143 pm px\textsuperscript{-1} across the spectral interval, and provides a resolving power of over R = 6 000. To obtain quantitative radiative data, the detector response was calibrated before each measurement campaign. A calibration lamp positioned at the meteorite sample location in the plasma wind tunnel was used to convert the digital camera output into absolute spectral radiance. Because the emitted radiance of the lamp was well characterized, the recorded counts could be reliably transformed into radiance units (W m\textsuperscript{-2} sr\textsuperscript{-1} nm\textsuperscript{-1}). During each ablation experiment, 19–127 frames of the meteorite emission were acquired, depending on the ablation duration and the selected camera gain. The average exposure time of one frame was 0.083 s. After calibration, the intensity profiles of the individual frames were summed to produce a representative spectrum for each meteorite. The relative line intensity ratios were verified to remain stable from frame to frame throughout the ablation experiment, justifying the summation approach and confirming that it does not bias the derived elemental abundances. Baseline subtraction had already been performed in the original dataset using the Fityk software \citep{Wojdyr:ko5121}, as described in our previous work in \citet{2024A&A...689A.323M}.

\subsection{Spectral data modeling and abundance determination}
\label{data_modeling}

To evaluate the radiating plasma conditions and compute elemental abundances, all calibrated and reduced spectra of ablated meteorites were fitted using the radiative transfer model introduced by \citet{1993A&A...279..627B}. This model considers the meteor plasma as a uniformly bright radiating gas cloud in local thermal equilibrium (LTE), incorporating self-absorption effects that modify the emission line profiles. The excitation state of atoms and ions follows the Boltzmann distribution, and the source function representing the ratio of emission to absorption is assumed to equal the Planck function describing black-body radiation at the plasma temperature. The intensity of an emission line depends on four key parameters: the plasma temperature $T$, the element column density $N$, the damping constant $\Gamma$, and the surface area of the radiating volume $P$. The damping constant controls the width of spectral lines in the model, representing the natural line broadening. The model employs a least-squares fitting method to invert the radiative transfer equation and derive these parameters from the fitted line intensities, from which relative elemental abundances are calculated.

\begin{figure}
	\centering
	\includegraphics[width=1\columnwidth]{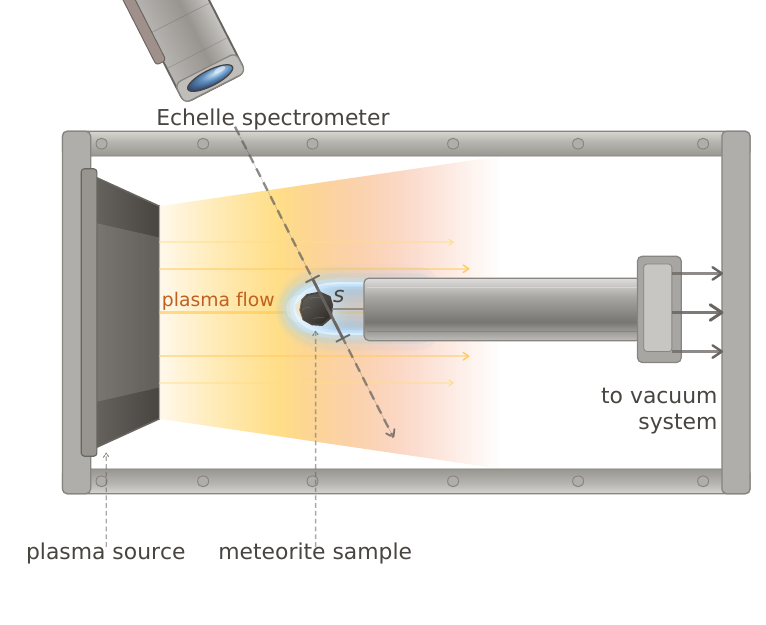}
	\caption{Top view schematic of the experimental geometry at the PWK1 plasma wind tunnel facility. The geometrical thickness $s$ denotes the extent of the radiating plasma along the line of sight as observed by the Echelle spectrometer.}
	\label{experiment}
\end{figure}

To obtain real chemical composition (elemental mass ratios), ionization corrections were applied. The ionization degree of each element is calculated using the Saha equation, allowing the combined contribution of neutral and ionized species to be accounted for. A simple geometrical model of the meteor head, with known surface area, allows the estimation of the free-electron density needed for ionization corrections, based on the calculated geometrical thickness $s$ of the radiating plasma along the line of sight. The column density of each element is converted to a volume density, and the electron density is then derived from the condition of neutrality using the Saha equation, accounting for the contributions of electrons from all elements. In meteor observations, the geometrical thickness of the radiating plasma is typically determined from the angular distance of the meteor from its radiant which is calculated from the meteor trajectory. For a more detailed description refer to \citet{1993A&A...279..627B}. In our laboratory experiments, the geometrical thickness was directly estimated from video recordings of individual meteorite ablation experiments. For simplicity and consistency, the same value of $s=7$ cm was applied to all meteorites in the dataset. A schematic illustration of the experimental geometry, including the definition of $s$, is provided in Fig. \ref{experiment}.

In our modeling procedure, only the low-temperature (main) component of the meteor spectrum was considered, as the majority of the emission originates from this regime. This choice reflects the conditions of our experiments, which simulated slow meteor events in which the high-temperature component is generally absent or negligible \citep{1994P&SS...42..145B}. Modeling the main spectrum allowed the determination of relative abundances for both major species commonly observed in meteor spectra (e.g., Fe I, Mg I, Cr I, Mn I, Si I, Na I, Ni I, Li I, and K I) and minor species such as Co I, Cu I, and V I for each sample. Emission from refractory elements (Al I, Ti I, and Ca I) was not detected \citep{2024A&A...689A.323M}, as discussed later.

The plasma parameters, specifically the temperature and column density of Fe I atoms, were first derived by fitting multiple Fe I lines. For the majority of samples, the fitting of iron lines was performed across the 400–560 nm wavelength range. In several cases, the observed spectrum below 450 nm appeared systematically underestimated with respect to the synthetic spectrum produced by the radiative transfer model. This behavior is attributed to increased uncertainty in the instrumental response calibration at short wavelengths, associated with the limited output of the integrating sphere in the near-UV region. In these cases, the fitting window for Fe I lines was restricted to the 450–560 nm region to ensure robustness. In all cases, special attention was paid to reliably reproducing the higher excitation Fe I multiplets, particularly multiplets of Fe I-318, Fe I-383, Fe I-553, and Fe I-686. The observed presence of these transitions is critical diagnostics, as they indicate higher excitation temperature and sufficiently high density to support the assumption of LTE in the plasma. Once these parameters were established, the abundances of the remaining elements relative to Fe were determined through careful fitting of the spectrum in the full wavelength range. Any spectral lines that clearly deviated from the LTE assumption or showed signs of saturation were excluded from the set of lines used for fitting. The estimated uncertainties in elemental abundances represent a combination of global and local uncertainties. First, we determined individual upper and lower limits based on the local quality of the spectral fit. Second, we evaluated the sensitivity of the derived abundances to variations in the adopted plasma parameters, resulting from the selection of the fitting wavelength range.

\section{Results}
\label{Results}

The radiative transfer modeling was applied to the set of 22 meteor ablation spectra presented in our previous study \citep{2024A&A...689A.323M}. In that study, we focused on identifying spectral features that can be individually resolved in meteor spectra and are therefore suitable for compositional diagnostics, in particular relative emission line intensity ratios. The analysis demonstrated that systematic variations in the relative intensities of Mg I, Fe I, Na I, Cr I, Mn I, Si I, H I, CN, Ni I, and Li I reflect differences in the original meteorite chemical composition and enable spectral discrimination between ordinary and carbonaceous chondrites, diverse achondrite types, stony-iron, and iron meteorites.

The observed spectral differences were shown to broadly trace bulk chemical composition variations, although, individual line intensity ratios do not always uniquely separate chemically similar meteorite classes \citep{2024A&A...689A.323M}. These limitations arise from intrinsic compositional overlaps between meteorite groups for specific element abundance ratios, sample heterogeneity within the same meteorite, and the fact that the composition of the radiating meteor plasma may not fully reproduce the bulk meteorite composition due to incomplete evaporation of refractory elements \citep{1993A&A...279..627B}. Furthermore, the observed line intensities are influenced by the physical conditions of the radiating plasma, with different line multiplets responding differently to temperature and self-absorption effects. As a result, analyses based primarily on relative line intensities cannot provide direct quantitative elemental abundances.

The approach presented in this work addresses these limitations by deriving quantitative physical parameters of the radiating plasma and elemental abundances through comprehensive radiative transfer modeling of laboratory-obtained meteor spectra. The analysis covers the most representative meteorite fall types, including ordinary, carbonaceous, and enstatite chondrites, as well as a wide range of achondrites (aubrite, lunar, martian, ureilite, howardite, eucrite, and diogenite), mesosiderite, and iron meteorites (list of all meteorites can be found in Table \ref{tab:plasma_conditions} and \ref{tab:abundances}).

\begin{figure*}[t]
	\centering
	\includegraphics[width=.9\textwidth]{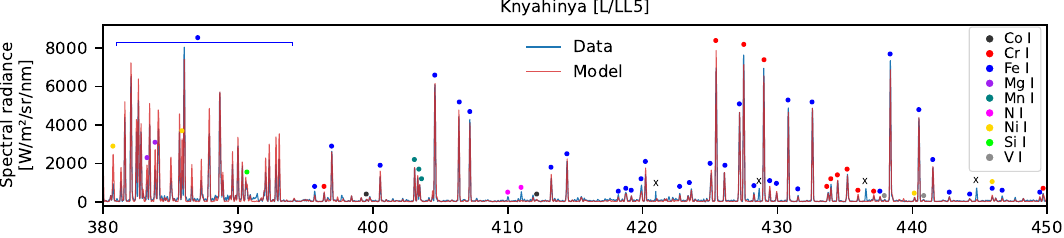}
    \centering
	\includegraphics[width=.9\textwidth]{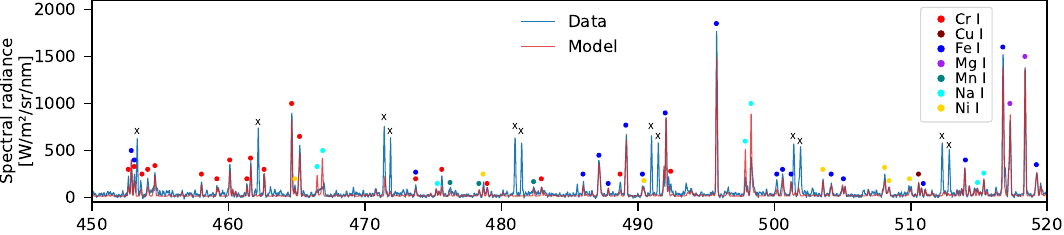}
    \centering
	\includegraphics[width=.9\textwidth]{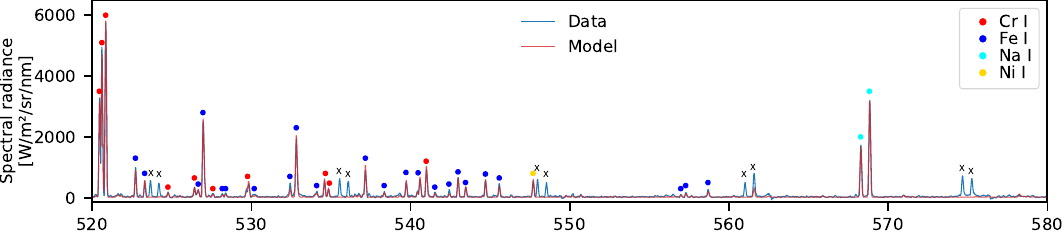}
    \centering
	\includegraphics[width=.9\textwidth]{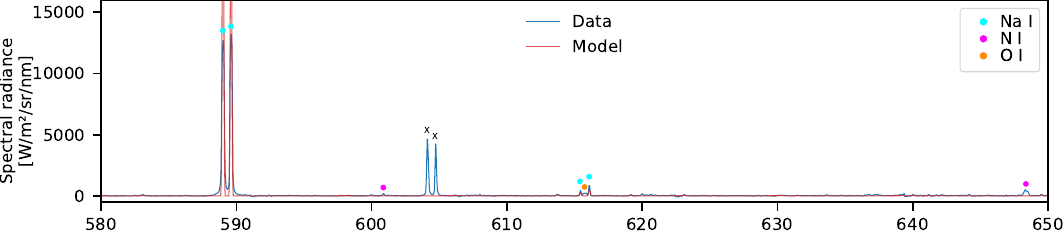}
    \centering
	\includegraphics[width=.9\textwidth]{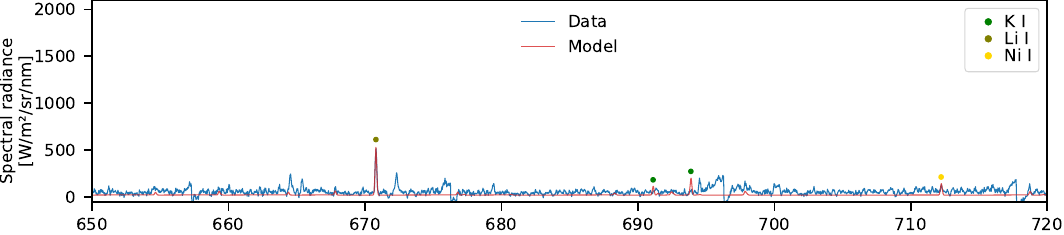}
    \centering
	\includegraphics[width=.9\textwidth]{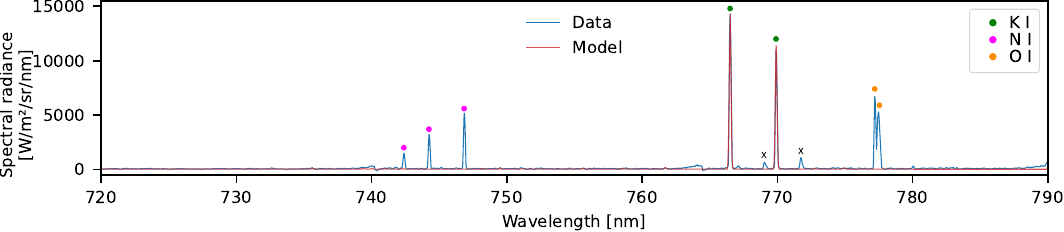}
	\caption{Observed high-resolution Echelle spectrum (blue) of the ordinary chondrite Knyahinya (L/LL5) compared with the final fit (red) obtained from the radiative transfer model assuming a single low-temperature component. The spectrum is divided into multiple wavelength ranges for better clarity. Colored markers indicate identified emission lines, while lines marked “X” correspond to artifacts from the reduction of Echelle spectra as reported in our previous work \citep{2024A&A...689A.323M}.}
	\label{KNY_profiles}
\end{figure*}

A direct comparison between the observed laboratory meteor spectra and the synthetic spectra generated by our model reveals a high degree of agreement across the analyzed wavelength range (380–780 nm) in most cases. Fig. \ref{KNY_profiles} shows the final fit for the ordinary chondrite Knyahinya (L/LL5). The model successfully reproduces the dominant emission features, confirming that the assumption of LTE with a single low-temperature component is valid under the plasma wind tunnel ablation conditions representative of low-speed entry meteor analogs. Discrepancies are generally confined to spectral regions with slightly poorer calibration (below $\sim$400\ nm), regions containing identified emission lines of O I and N I originating from the plasma flow and H I emission associated with the high-temperature spectral component that was not considered in the modeling, and the CN molecular band region ($\sim$384-389\ nm). Both the H I and CN emissions, which are regarded as markers of water and organic compounds in meteoroids, were analyzed in detail in our previous studies \citep{2023Icar..40415682P, 2024A&A...689A.323M}. These features do not affect the derivation of abundances for the atomic species analyzed in this work.

\clearpage
\clearpage

\begin{sidewaystable*}[p]
\centering
\caption{Overview of the meteorites analyzed in this study, including their name, meteorite type, and abbreviation used in the figures. The table lists the parameters from modeling the main spectral component using the radiative transfer model, namely the temperature (T), Fe I column density (N\textsubscript{Fe I}), damping constant ($\Gamma$), and radiating plasma cross section (P), as well as the derived relative abundances (column densities) of selected elements with respect to Fe I (before ionization correction). Abundances of V and Co represent upper limits, as only a few of their spectral lines were visible in the spectra and these were generally close to the noise level.}
\label{tab:plasma_conditions}
\scriptsize
\setlength{\tabcolsep}{3pt} 
\begin{tabular}{@{}lllrrrrrrrrrrrrrrrr@{}}
\toprule\toprule
\textbf{Meteorite} & \textbf{Type} & \textbf{Abbreviation} 
& \textbf{T {[}K{]}} & \textbf{N\textsubscript{Fe I} {[}cm\textsuperscript{-2}{]}} & \textbf{$\Gamma$ {[}s\textsuperscript{-1}{]}} & \textbf{P {[}m\textsuperscript{2}{]}} & \textbf{Si I} & \textbf{Mg I} & \textbf{Fe I} & \textbf{Cr I} & \textbf{Mn I} & \textbf{Na I} & \textbf{K I} & \textbf{Li I} & \textbf{Ni I} & \textbf{V I} & \textbf{Co I} & \textbf{Cu I} \\ 
\specialrule{0.1em}{0pt}{0pt}\addlinespace[0.3em]\specialrule{0.1em}{0pt}{0pt}
Buzzard Coulee & H4 & BUZ{[}H4{]} & 5470 & $1.3 \times 10^{15}$ & $6.0 \times 10^{9}$ & 8.7 & 1.10 & 0.061 & 1.00 & 0.009 & 0.006 & 0.052 & 0.007 & 0.00003 & 0.15 & 0.00010 & 0.003 & 0.008 \\
 &  &  & ± 190 & ± $8.8 \times 10^{14}$ & ± $5.0 \times 10^{8}$ & ± 10 & ± 0.05 & ± 0.003 & - & ± 0.00001 & ± 0.0004 & ± 0.002 & ± 0.004 & ± 0.00001 & ± 0.02 & - & - & ± 0.002 \\
Košice & H5 & KOS{[}H5{]} & 5640 & $1.2 \times 10^{14}$ & 0.0 & 18.9 & 0.83 & 0.052 & 1.00 & 0.003 & 0.005 & 0.099 & 0.005 & 0.00005 & 0.26 & 0.00020 & 0.009 & 0.010 \\
 &  &  & ± 160 & ± $8.5 \times 10^{13}$ & ± 0.0 & ± 26.9 & ± 0.08 & ± 0.002 & - & ± 0.001 & ± 0.001 & ± 0.002 & ± 0.001 & ± 0.00001 & ± 0.03 & - & - & ± 0.0004 \\
Pultusk & H5 & PUL{[}H5{]} & 5220 & $3.8 \times 10^{15}$ & $1.1 \times 10^{9}$ & 9.0 & 1.05 & 0.054 & 1.00 & 0.006 & 0.006 & 0.034 & 0.006 & 0.00002 & 0.14 & 0.00005 & 0.002 & 0.010 \\
 &  &  & ± 90 & ± $1.4 \times 10^{15}$ & ± $3.4 \times 10^{9}$ & ± 2.8 & ± 0.05 & ± 0.003 & - & ± 0.0004 & ± 0.002 & ± 0.001 & ± 0.003 & ± 0.00000 & ± 0.01 & - & - & ± 0.001 \\
NWA 869 & L3-6 & NWA{[}L3-6{]} & 5490 & $6.8 \times 10^{13}$ & $1.0 \times 10^{9}$ & 77.8 & 0.90 & 0.058 & 1.00 & 0.006 & 0.008 & 0.137 & 0.010 & 0.00009 & 0.24 & 0.00075 & 0.010 & 0.062 \\
 &  &  & ± 210 & ± $6.5 \times 10^{14}$ & ±$1.5 \times 10^{10}$ & ± 66.9 & ± 0.20 & ± 0.006 & - & ± 0.003 & ± 0.004 & ± 0.010 & ± 0.033 & ± 0.00002 & ± 0.004 & - & - & ± 0.007 \\
Mocs & L5-6 & MOC{[}L5-6{]} & 5600 & $1.1 \times 10^{14}$ & $1.0 \times 10^{10}$ & 10.4 & 1.15 & 0.064 & 1.00 & 0.005 & 0.006 & 0.120 & 0.017 & 0.00007 & 0.36 & 0.00066 & 0.019 & 0.012 \\
 &  &  & ± 60 & ± $2.2 \times 10^{14}$ & ± 0.0 & ± 6.7 & ± 0.05 & ± 0.009 & - & ± 0.001 & ± 0.002 & ± 0.010 & ± 0.043 & ± 0.00001 & ± 0.07 & - & - & ± 0.001 \\
Knyahinya & L/LL5 & KNY{[}L/LL5{]} & 5310 & $4.4 \times 10^{14}$ & $3.6 \times 10^{9}$ & 7.0 & 1.50 & 0.072 & 1.00 & 0.050 & 0.006 & 0.158 & 0.017 & 0.00006 & 0.65 & 0.00020 & 0.008 & 0.021 \\
 &  &  & ± 220 & ± $5.4 \times 10^{13}$ & ± $4.1 \times 10^{8}$ & ± 1.1 & ± 0.20 & ± 0.010 & - & ± 0.001 & ± 0.001 & ± 0.011 & ± 0.002 & ± 0.00001 & ± 0.05 & - & - & ± 0.0004 \\
Ragland & LL3.4 & RAG{[}LL3.4{]} & 5560 & $1.0 \times 10^{14}$ & $2.3 \times 10^{9}$ & 7.3 & 0.93 & 0.063 & 1.00 & 0.256 & 0.005 & 0.285 & 0.051 & 0.00056 & 1.70 & 0.00080 & 0.028 & 0.498 \\
 &  &  & ± 380 & ± $1.0 \times 10^{13}$ & ± $3.1 \times 10^{9}$ & ± 2.6 & ± 0.20 & ± 0.019 & - & ± 0.026 & ± 0.002 & ± 0.055 & ± 0.001 & ± 0.00007 & ± 0.43 & - & - & ± 0.073 \\
Chelyabinsk & LL5 & CHE{[}LL5{]} & 5410 & $7.9 \times 10^{14}$ & $9.0 \times 10^{9}$ & 7.7 & 1.13 & 0.075 & 1.00 & 0.012 & 0.005 & 0.123 & 0.009 & 0.00005 & 0.49 & 0.00014 & 0.007 & 0.013 \\
 &  &  & ± 250 & ± $5.6 \times 10^{14}$ & ± $2.0 \times 10^{9}$ & ± 9.8 & ± 0.02 & ± 0.010 & - & ± 0.002 & ± 0.001 & ± 0.009 & ± 0.005 & ± 0.00001 & ± 0.01 & - & - & ± 0.001 \\
Kheneg Ljouâd & LL5/6 & KLJ{[}LL5/6{]} & 5560 & $1.3 \times 10^{14}$ & $7.0 \times 10^{9}$ & 5.2 & 0.80 & 0.062 & 1.00 & 0.016 & 0.007 & 0.354 & 0.120 & 0.00014 & 0.80 & 0.00090 & 0.020 & 0.039 \\
 &  &  & ± 370 & ± $2.7 \times 10^{14}$ & ± 0.0 & ± 2.5 & ± 0.20 & ± 0.026 & - & ± 0.003 & ± 0.004 & ± 0.159 & ± 0.020 & ± 0.00002 & ± 0.16 & - & - & ± 0.004 \\
Eagle & EH5 & EAG{[}EH5{]} & 5570 & $1.9 \times 10^{14}$ & 0.0 & 11.1 & 1.50 & 0.048 & 1.00 & 0.003 & 0.007 & 0.065 & 0.011 & 0.00004 & 0.14 & 0.00010 & 0.003 & 0.009 \\
 &  &  & ± 140 & ± $1.9 \times 10^{13}$ & ± 0.0 & ± 2.3 & ± 0.10 & ± 0.006 & - & ± 0.001 & ± 0.002 & ± 0.004 & ± 0.010 & ± 0.000002 & ± 0.01 & - & - & ± 0.0004 \\
Allende & CV3 & ALL{[}CV3{]} & 5390 & $2.8 \times 10^{14}$ & 0.0 & 14.6 & 0.37 & 0.066 & 1.00 & 0.004 & 0.003 & 0.111 & 0.008 & 0.00006 & 0.73 & 0.00015 & 0.012 & 0.056 \\
 &  &  & ± 100 & ± $9.0 \times 10^{12}$ & ± 0.0 & ± 1.4 & ± 0.05 & ± 0.004 & - & ± 0.0002 & ± 0.001 & ± 0.003 & ± 0.002 & ± 0.000004 & ± 0.02 & - & - & ± 0.0003 \\
Lancé & CO3.5 & LAN{[}CO3.5{]} & 5770 & $6.8 \times 10^{13}$ & $1.0 \times 10^{10}$ & 9.8 & 0.46 & 0.054 & 1.00 & 0.004 & 0.005 & 0.131 & 0.018 & 0.00009 & 0.32 & 0.00048 & 0.048 & 0.038 \\
 &  &  & ± 300 & ± $1.1 \times 10^{14}$ & ± 0.0 & ± 4.6 & ± 0.10 & ± 0.013 & - & ± 0.001 & ± 0.001 & ± 0.017 & ± 0.012 & ± 0.00002 & ± 0.03 & - & - & ± 0.002 \\
Murchison & CM2 & MUR{[}CM2{]} & 5790 & $3.8 \times 10^{14}$ & $8.0 \times 10^{9}$ & 0.6 & 0.73 & 0.035 & 1.00 & 0.007 & 0.013 & 0.233 & 0.100 & 0.00024 & 0.39 & 0.00025 & 0.016 & 0.090 \\
 &  &  & ± 540 & ± $1.7 \times 10^{14}$ & ± $1.0 \times 10^{9}$ & ± 0.1 & ± 0.27 & ± 0.024 & - & ± 0.002 & ± 0.005 & ± 0.062 & ± 0.020 & ± 0.00005 & ± 0.09 & - & - & ± 0.015 \\
Sariçiçek & Howardite & SAR{[}HOW{]} & 5540 & $1.3 \times 10^{14}$ & 0.0 & 5.8 & 4.35 & 0.050 & 1.00 & 0.058 & 0.009 & 0.134 & 0.005 & 0.00029 & 0.07 & 0.00024 & 0.006 & 0.029 \\
 &  &  & ± 420 & ± $2.3 \times 10^{13}$ & ± 0.0 & ± 4.9 & ± 0.60 & ± 0.031 & - & ± 0.013 & ± 0.004 & ± 0.042 & ± 0.002 & ± 0.00004 & ± 0.03 & - & - & ± 0.005 \\
Stannern & Eucrite & STA{[}EUC{]} & 5760 & $1.6 \times 10^{14}$ & $8.5 \times 10^{9}$ & 4.3 & 3.86 & 0.023 & 1.00 & 0.026 & 0.010 & 0.151 & 0.003 & 0.00022 & 0.03 & 0.00022 & 0.012 & 0.002 \\
 &  &  & ± 350 & ± $6.3 \times 10^{14}$ & ± $5.0 \times 10^{8}$ & ± 2.7 & ± 0.50 & ± 0.001 & - & ± 0.002 & ± 0.005 & ± 0.012 & ± 0.004 & ± 0.00007 & ± 0.00 & - & - & ± 0.000 \\
Bilanga & Diogenite & BIL{[}DIO{]} & 5510 & $1.8 \times 10^{14}$ & $7.0 \times 10^{9}$ & 8.0 & 4.65 & 0.121 & 1.00 & 0.259 & 0.009 & 0.009 & 0.005 & 0.00037 & 0.02 & 0.00083 & 0.009 & 0.085 \\
 &  &  & ± 320 & ± $6.3 \times 10^{13}$ & ± $1.0 \times 10^{9}$ & ± 5.4 & ± 1.95 & ± 0.101 & - & ± 0.054 & ± 0.003 & ± 0.006 & ± 0.001 & ± 0.00006 & ± 0.00 & - & - & ± 0.041 \\
Tissint & Shergottite & TIS{[}SHE{]} & 5790 & $2.3 \times 10^{13}$ & $6.0 \times 10^{9}$ & 11.7 & 2.93 & 0.049 & 1.00 & 0.057 & 0.012 & 0.281 & 0.010 & 0.00027 & 0.15 & 0.00127 & 0.024 & 0.015 \\
 &  &  & ± 130 & ± $8.0 \times 10^{13}$ & ± $1.0 \times 10^{9}$ & ± 8.6 & ± 0.10 & ± 0.004 & - & ± 0.004 & ± 0.003 & ± 0.021 & ± 0.015 & ± 0.00002 & ± 0.002 & - & - & ± 0.002 \\
NWA 11303 & Lunar & LUN{[}LUN{]} & 5470 & $1.2 \times 10^{14}$ & $1.0 \times 10^{10}$ & 2.4 & 10.04 & 0.123 & 1.00 & 0.016 & 0.004 & 0.204 & 0.072 & 0.00053 & 0.25 & 0.00179 & 0.020 & 0.022 \\
 &  &  & ± 300 & ± $2.5 \times 10^{14}$ & ± $1.0 \times 10^{9}$ & ± 1.7 & ± 0.50 & ± 0.036 & - & ± 0.005 & ± 0.002 & ± 0.034 & ± 0.013 & ± 0.00002 & ± 0.05 & - & - & ± 0.010 \\
Norton County & Aubrite & NCO{[}AUB{]} & 5530 & $8.7 \times 10^{13}$ & $9.0 \times 10^{9}$ & 9.5 & 25.00 & 1.966 & 1.00 & 0.026 & 0.068 & 0.329 & 0.038 & 0.00022 & 0.17 & 0.00064 & 0.016 & 0.032 \\
 &  &  & ± 320 & ± $1.0 \times 10^{13}$ & ± $2.8 \times 10^{9}$ & ± 1.0 & ± 15.00 & ± 1.253 & - & ± 0.007 & ± 0.023 & ± 0.129 & ± 0.003 & ± 0.000001 & ± 0.06 & - & - & ± 0.010 \\
Dhofar 1575 & Ureilite & DHO{[}URE{]} & 5690 & $2.9 \times 10^{14}$ & 0.0 & 0.6 & 0.80 & 0.147 & 1.00 & 0.021 & 0.011 & 0.009 & 0.001 & 0.00032 & 0.12 & 0.00027 & 0.022 & 0.063 \\
 &  &  & ± 540 & ± $1.7 \times 10^{14}$ & ± 0.0 & ± 0.1 & ± 0.25 & ± 0.103 & - & ± 0.005 & ± 0.008 & ± 0.003 & ± 0.0002 & ± 0.00007 & ± 0.04 & - & - & ± 0.008 \\
Mincy & Mesosiderite & MIN{[}MES{]} & 5810 & $1.1 \times 10^{14}$ & 0.0 & 4.0 & 1.43 & 0.042 & 1.00 & 0.028 & 0.006 & 0.023 & 0.003 & 0.00017 & 0.40 & 0.00122 & 0.027 & 0.016 \\
 &  &  & ± 110 & ± $1.1 \times 10^{14}$ & ± 0.0 & ± 2.3 & ± 0.10 & ± 0.003 & - & ± 0.003 & ± 0.002 & ± 0.001 & ± 0.003 & ± 0.00001 & ± 0.08 & - & - & ± 0.001 \\
Mount Joy & Iron & MJO{[}IRON{]} & 5360 & $4.0 \times 10^{14}$ & 0.0 & 3.6 & 0.05 & 0.001 & 1.00 & 0.0001 & 0.0003 & 0.001 & 0.00005 & 0.00001 & 0.06 & 0.00016 & 0.005 & 0.070 \\
 &  &  & ± 250 & ± $2.4 \times 10^{14}$ & ± 0.0 & ± 2.7 & ± 0.01 & ± 0.00003 & - & ± 0.00004 & ± 0.00027 & ± 0.0001 & ± 0.00001 & ± 0.000002 & ± 0.01 & - & - & ± 0.001 \\ 
\bottomrule
\end{tabular}
\end{sidewaystable*}

\begin{sidewaystable*}[p]
\centering
\caption{Final elemental abundances of the analyzed meteorites after ionization correction, expressed as elemental mass fractions relative to Fe. The table includes the meteorite name, meteorite type, recovery type, and the calculated electron density, followed by the derived elemental abundances. Elemental mass fractions of V and Co represent upper limits.}
\label{tab:abundances}
\scriptsize
\setlength{\tabcolsep}{3pt} 
\begin{tabular}{@{}llcrrrrrrrrrrrrr@{}}
\toprule\toprule
\textbf{Meteorite} & \textbf{Type} & \textbf{Rec. T.} 
& $\boldsymbol{n_e\,[cm^{-3}]}$ & \textbf{Si} & \textbf{Mg} & \textbf{Fe} & \textbf{Cr} & \textbf{Mn} & \textbf{Na} & \textbf{K} & \textbf{Li} & \textbf{Ni} & \textbf{V} & \textbf{Co} & \textbf{Cu} \\
\specialrule{0.1em}{0pt}{0pt}\addlinespace[0.3em]\specialrule{0.1em}{0pt}{0pt}
Buzzard Coulee & H4 & fall & $5.63 \times 10^{14}$ & 0.46 & 0.033 & 1.00 & 0.015 & 0.007 & 0.530 & 0.60 & 0.00005 & 0.14 & 0.00021 & 0.003 & 0.008 \\
 &  &  &  & ± 0.08 & ± 0.005 & - & ± 0.006 & ± 0.001 & ± 0.390 & ± 0.22 & ± 0.00006 & ± 0.01 & - & - & ± 0.00004 \\
Košice & H5 & fall & $2.45 \times 10^{14}$ & 0.25 & 0.036 & 1.00 & 0.008 & 0.008 & 1.953 & 0.80 & 0.00018 & 0.19 & 0.00070 & 0.008 & 0.008 \\
 &  &  &  & ± 0.08 & ± 0.003 & - & ± 0.003 & ± 0.0003 & ± 0.502 & ± 0.16 & ± 0.00007 & ± 0.01 & - & - & ± 0.001 \\
Pultusk & H5 & fall & $6.38 \times 10^{14}$ & 0.49 & 0.026 & 1.00 & 0.008 & 0.007 & 0.206 & 0.33 & 0.00002 & 0.13 & 0.00007 & 0.002 & 0.010 \\
 &  &  &  & ± 0.03 & ± 0.001 & - & ± 0.0003 & ± 0.002 & ± 0.020 & ± 0.17 & ± 0.000001 & ± 0.01 & - & - & ± 0.001 \\
NWA 869 & L3-6 & find & $1.86 \times 10^{14}$ & 0.29 & 0.040 & 1.00 & 0.017 & 0.012 & 2.870 & 1.73 & 0.00033 & 0.18 & 0.00255 & 0.009 & 0.051 \\
 &  &  &  & ± 0.22 & ± 0.008 & - & ± 0.012 & ± 0.008 & ± 1.921 & ± 0.65 & ± 0.00025 & ± 0.05 & - & - & ± 0.009 \\
Mocs & L5-6 & fall & $2.82 \times 10^{14}$ & 0.37 & 0.043 & 1.00 & 0.013 & 0.009 & 2.179 & 2.37 & 0.00022 & 0.28 & 0.00212 & 0.016 & 0.010 \\
 &  &  &  & ± 0.10 & ± 0.011 & - & ± 0.002 & ± 0.001 & ± 0.989 & ± 2.55 & ± 0.00010 & ± 0.03 & - & - & ± 0.0001 \\
Knyahinya & L/LL5 & fall & $4.36 \times 10^{14}$ & 0.66 & 0.038 & 1.00 & 0.080 & 0.007 & 1.512 & 1.35 & 0.00010 & 0.61 & 0.00039 & 0.008 & 0.021 \\
 &  &  &  & ± 0.14 & ± 0.002 & - & ± 0.014 & ± 0.002 & ± 0.232 & ± 0.39 & ± 0.00004 & ± 0.09 & - & - & ± 0.002 \\
Ragland & LL3.4 & find & $4.10 \times 10^{14}$ & 0.34 & 0.038 & 1.00 & 0.536 & 0.007 & 4.015 & 5.57 & 0.00142 & 1.43 & 0.00211 & 0.026 & 0.451 \\
 &  &  &  & ± 0.13 & ± 0.008 & - & ± 0.021 & ± 0.004 & ± 0.706 & ± 0.84 & ± 0.00032 & ± 0.52 & - & - & ± 0.124 \\
Chelyabinsk & LL5 & fall & $5.49 \times 10^{14}$ & 0.49 & 0.040 & 1.00 & 0.020 & 0.007 & 1.149 & 0.66 & 0.00008 & 0.46 & 2.73822 & 0.007 & 0.013 \\
 &  &  &  & ± 0.11 & ± 0.003 & - & ± 0.014 & ± 0.003 & ± 0.744 & ± 0.21 & ± 0.00010 & ± 0.09 & - & - & ± 0.002 \\
Kheneg Ljouâd & LL5/6 & fall & $6.05 \times 10^{14}$ & 0.32 & 0.035 & 1.00 & 0.030 & 0.008 & 3.806 & 10.03 & 0.00027 & 0.72 & 0.00199 & 0.019 & 0.038 \\
 &  &  &  & ± 0.15 & ± 0.007 & - & ± 0.013 & ± 0.005 & ± 1.348 & ± 6.07 & ± 0.00017 & ± 0.25 & - & - & ± 0.010 \\
Eagle & EH5 & fall & $2.70 \times 10^{14}$ & 0.50 & 0.032 & 1.00 & 0.008 & 0.011 & 1.171 & 1.49 & 0.00011 & 0.11 & 0.00030 & 0.003 & 0.007 \\
 &  &  &  & ± 0.02 & ± 0.003 & - & ± 0.002 & ± 0.004 & ± 0.187 & ± 0.99 & ± 0.000002 & ± 0.01 & - & - & ± 0.001 \\
Allende & CV3 & fall & $3.10 \times 10^{14}$ & 0.15 & 0.039 & 1.00 & 0.008 & 0.004 & 1.565 & 0.90 & 0.00015 & 0.64 & 0.00036 & 0.012 & 0.054 \\
 &  &  &  & ± 0.01 & ± 0.001 & - & ± 0.00002 & ± 0.002 & ± 0.199 & ± 0.21 & ± 0.00002 & ± 0.03 & - & - & ± 0.002 \\
Lancé & CO3.5 & fall & $2.64 \times 10^{14}$ & 0.12 & 0.039 & 1.00 & 0.011 & 0.007 & 2.612 & 2.60 & 0.00033 & 0.22 & 0.00178 & 0.039 & 0.028 \\
 &  &  &  & ± 0.09 & ± 0.00003 & - & ± 0.005 & ± 0.003 & ± 0.648 & ± 0.75 & ± 0.00015 & ± 0.08 & - & - & ± 0.006 \\
Murchison & CM2 & fall & $1.10 \times 10^{15}$ & 0.29 & 0.020 & 1.00 & 0.012 & 0.016 & 2.142 & 6.52 & 0.00041 & 0.34 & 0.00055 & 0.015 & 0.085 \\
 &  &  &  & ± 0.18 & ± 0.009 & - & ± 0.006 & ± 0.008 & ± 0.559 & ± 3.15 & ± 0.00023 & ± 0.14 & - & - & ± 0.028 \\
Sariçiçek & Howardite & fall & $2.62 \times 10^{14}$ & 1.47 & 0.033 & 1.00 & 0.139 & 0.013 & 2.417 & 0.78 & 0.00091 & 0.05 & 0.00075 & 0.005 & 0.025 \\
 &  &  &  & ± 0.08 & ± 0.013 & - & ± 0.052 & ± 0.006 & ± 0.508 & ± 0.33 & ± 0.00025 & ± 0.04 & - & - & ± 0.009 \\
Stannern & Eucrite & fall & $3.73 \times 10^{14}$ & 1.18 & 0.016 & 1.00 & 0.064 & 0.014 & 2.584 & 0.31 & 0.00068 & 0.02 & 0.00072 & 0.011 & 0.002 \\
 &  &  &  & ± 0.28 & ± 0.003 & - & ± 0.027 & ± 0.009 & ± 1.195 & ± 0.17 & ± 0.00045 & ± 0.01 & - & - & ± 0.0005 \\
Bilanga & Diogenite & fall & $1.80 \times 10^{14}$ & 1.44 & 0.083 & 1.00 & 0.692 & 0.014 & 0.194 & 0.93 & 0.00141 & 0.01 & 0.00292 & 0.008 & 0.068 \\
 &  &  &  & ± 0.94 & ± 0.057 & - & ± 0.144 & ± 0.007 & ± 0.184 & ± 0.09 & ± 0.00028 & ± 0.002 & - & - & ± 0.045 \\
Tissint & Shergottite & fall & $1.98 \times 10^{14}$ & 0.70 & 0.038 & 1.00 & 0.169 & 0.020 & 6.209 & 1.60 & 0.00109 & 0.10 & 0.00510 & 0.019 & 1.012 \\
 &  &  &  & ± 0.34 & ± 0.004 & - & ± 0.032 & ± 0.007 & ± 1.610 & ± 1.33 & ± 0.00040 & ± 0.03 & - & - & ± 0.997 \\
NWA 11303 & Lunar & find & $4.12 \times 10^{14}$ & 3.98 & 0.071 & 1.00 & 0.030 & 0.005 & 2.556 & 7.29 & 0.00117 & 0.22 & 0.00423 & 0.019 & 0.021 \\
 &  &  &  & ± 0.15 & ± 0.020 & - & ± 0.010 & ± 0.002 & ± 0.926 & ± 1.41 & ± 0.00020 & ± 0.05 & - & - & ± 0.010 \\
Norton County & Aubrite & fall & $3.85 \times 10^{14}$ & 9.39 & 1.185 & 1.00 & 0.055 & 0.091 & 4.654 & 4.22 & 0.00056 & 0.15 & 0.00167 & 0.014 & 0.029 \\
 &  &  &  & ± 8.09 & ± 0.527 & - & ± 0.024 & ± 0.038 & ± 0.563 & ± 1.08 & ± 0.00015 & ± 0.07 & - & - & ± 0.013 \\
Dhofar 1575 & Ureilite & find & $1.94 \times 10^{14}$ & 0.21 & 0.108 & 1.00 & 0.060 & 0.018 & 0.205 & 0.19 & 0.00127 & 0.08 & 0.00106 & 0.017 & 0.045 \\
 &  &  &  & ± 0.20 & ± 0.040 & - & ± 0.026 & ± 0.014 & ± 0.032 & ± 0.04 & ± 0.00047 & ± 0.06 & - & - & ± 0.022 \\
Mincy & Mesosiderite & find & $1.76 \times 10^{14}$ & 0.32 & 0.033 & 1.00 & 0.086 & 0.009 & 0.527 & 0.53 & 0.00071 & 0.25 & 0.00506 & 0.021 & 0.010 \\
 &  &  &  & ± 0.06 & ± 0.004 & - & ± 0.005 & ± 0.003 & ± 0.108 & ± 0.29 & ± 0.00009 & ± 0.07 & - & - & ± 0.00001 \\
Mount Joy & Iron & find & $8.78 \times 10^{13}$ & 0.01 & 0.001 & 1.00 & 0.0003 & 0.0004 & 0.036 & 0.01 & 0.00005 & 0.04 & 0.00065 & 0.004 & 0.054 \\
 &  &  &  & ± 0.006 & ± 0.0001 & - & ± 0.0001 & ± 0.0001 & ± 0.008 & ± 0.002 & ± 0.00001 & ± 0.01 & - & - & ± 0.011 \\ 
\bottomrule
\end{tabular}
\end{sidewaystable*}

\subsection{Derived plasma parameters}

Using the fitting strategy described in Section \ref{data_modeling}, we derived the plasma conditions for all the tested meteorite samples. The resulting plasma temperature ($T$), the column density of neutral iron ($N\textsubscript{Fe I}$), the damping constant ($\Gamma$), and the cross section of the radiating volume ($P$), along with the relative elemental abundances (uncorrected for ionization) for all species identified in the main spectral component, are summarized in Table \ref{tab:plasma_conditions}.

Our modeling resulted in plasma temperatures in the range of 5220–5810 K, with an average estimated uncertainty of  $\pm$ 260 K. These values are notably higher than the estimates presented in our study in \citet{2024A&A...689A.323M}, where temperatures ranged from 3680 to 4830 K ($\pm$ 200 K). This discrepancy is likely primarily methodological. First and foremost, in our previous work we fitted most Fe I lines in the 406–546 nm range using a simplified LTE model without accounting for self-absorption, including the spectral region below 450 nm which, as discussed in Section 2.2, may introduce a systematic bias in the derived temperatures due to minor uncertainties in the instrument response calibration. Second, while the previous work relied solely on Fe I lines for temperature determination, the present work initially constrained the temperature by fitting Fe I lines and subsequently refined it by progressively including lines of additional elements. Additionally, by accounting for self-absorption in the radiative transfer model used in this work, we were able to include the strongest lines in the fit, whereas in the previous work these lines were excluded to avoid biasing the fit under the optically thin assumption. Together, these methodological improvements result in temperatures that more accurately reflect the excitation state of the dense plasma.

The derived column densities of Fe I atoms span nearly two orders of magnitude, ranging from \num{2.3e13}cm\textsuperscript{-2} to \num{3.8e15}cm\textsuperscript{-2}, with an average value of \num{4.2e14}cm\textsuperscript{-2}. As shown in Table 1, the emission in the majority of cases is consistent with an optically thick plasma rather than an optically thin approximation. This is supported by the high Fe I column densities and by the need to include a non-zero damping constant in order to achieve a reliable fit to the strongest optically thick emission lines. The derived damping constants have an average value of \num{4.5e9}s\textsuperscript{-1}. However, in some cases this parameter could only be estimated approximately due to the limited number of sufficiently bright lines approaching the optically thick limit.

The validity of our results can be further assessed by comparison with the numerical simulation of \citet{2024Icar..40715768P}, who used the DSMC (Direct Simulation Monte Carlo) method bi-directionally coupled with a radiation solver to simulate the atmospheric entry of the iron meteorite, with results showing good agreement with the experimentally measured spectrum of the iron Mount Joy (MJO) meteorite obtained in the plasma wind tunnel. For the MJO meteorite, our derived plasma temperature of T = 5360 $\pm$ 250 K is in reasonable agreement with the electronic excitation temperature of Fe of approximately 4500 K reported by \citet{2024Icar..40715768P}. Furthermore, converting our derived total Fe column density to a number density yields $n\textsubscript{Fe}$ $\approx$ \num{1.3e20}m\textsuperscript{-3}, which is in good agreement with the Fe number densities of the order \num{1e19} to \num{1e20}m\textsuperscript{-3} in the emitting region as obtained from the simulation of \citet{2024Icar..40715768P}. This consistency supports the validity of our LTE approach.

A comparison with bolide spectra reveals some physical differences. The plasma temperatures obtained from the modeling are slightly higher than the value of $\approx$\ 4500 K characteristic of the main spectral component of bolides modeled using the same approach \citep{1993A&A...279..627B, 1994P&SS...42..145B}. Conversely, the Fe I column densities determined in this work are generally lower than the typical bolide value of $\approx$\ \num{e16}cm\textsuperscript{-2}, although the majority exceeds the approximate limit for optically thin meteor radiation (\num{e13}cm\textsuperscript{-2}). The estimated average value of damping constant is in agreement with typical values for bolides ($\approx$ \num{4e9}s\textsuperscript{-1}). These results, particularly the combination of lower column densities and comparable or slightly higher damping constants, indicate the presence of a physically small but dense radiating plasma volume achieved in laboratory conditions.

The higher plasma temperatures observed in this study can likely be attributed to the specific laboratory simulation conditions, including energy input. The high-enthalpy flow conditions in the wind tunnel may maintain a higher excitation state compared to the typical conditions in a meteor plasma. Regarding the variations among the analyzed samples, the observed differences in determined temperatures and column densities of Fe I likely reflect the diversity in sample size and bulk composition. These factors directly influence the efficiency of ablation and the resulting density of the radiating vapor cloud.

\subsection{Elemental abundances and comparison with bulk composition}

Relative abundances determined from the radiative transfer model (Table \ref{tab:plasma_conditions}) were corrected for ionization using the Saha equation and converted from numbers of atoms to mass fractions. They are presented in Table \ref{tab:abundances} as final elemental mass fractions relative to Fe, together with the calculated electron density, whose average value was \num{3.7e14}cm\textsuperscript{-3}. As demonstrated by \citet{2004EM&P...95..245B}, accounting for the ionization degree of individual elements is critical for converting observed spectral line intensities in radiance units into real chemical composition, as a significant fraction of atoms in the meteor plasma exists in ionized states in both spectral components. Neglecting this correction would lead to biased elemental abundances due to differences in ionization potentials among the detected species, with K, Na, and Ca being the most strongly affected.

To evaluate the accuracy of our spectral modeling, we compared the calculated elemental abundances with the bulk composition of the investigated meteorites reported in the literature. The compilation of bulk compositions for all meteorite samples was taken from our previous work (\citet{2024A&A...689A.323M}, Table 1), and the full listing of reference values with literature sources is therefore not repeated here. Bulk composition data are not available for the Buzzard Coulee (H4) and Kheneg Ljouâd (LL5/6) meteorites. As we reported in \citet{2024A&A...689A.323M}, the mean compositions of 24 feldspathic lunar meteorites and six ureilites were adopted as representative proxies for the lunar and ureilite samples, respectively. Bulk abundance data for Li and minor species detected in our spectra of ablated meteorites, namely V, Co, and Cu, are rarely reported in the literature, and comparison for these elements was therefore not possible.

\begin{figure}
	\centering
	\includegraphics[width=1\columnwidth]{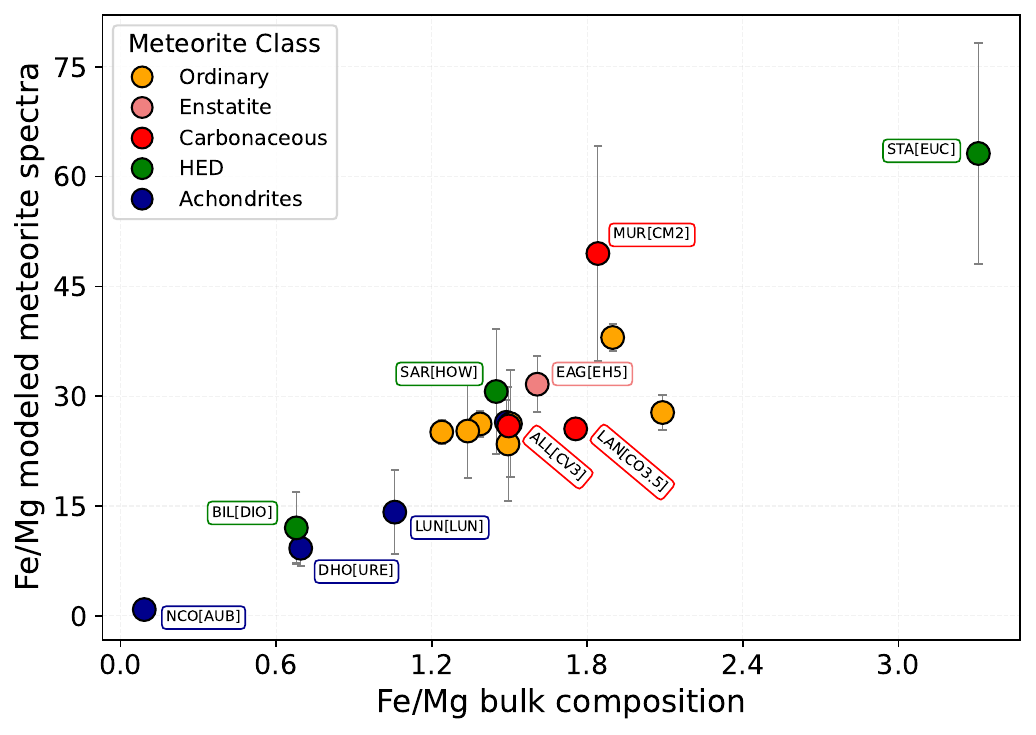}
	\caption{Correlation between the calculated Fe/Mg mass ratios derived from spectral modeling and literature bulk values for the tested meteorites. Reference bulk composition data are adopted from our previous study \citep{2024A&A...689A.323M}. Different colors correspond to different meteorite classes. The HED class refers to Howardite–Eucrite–Diogenite achondrites.}
	\label{Fe_to_Mg}
\end{figure}

\begin{figure}
	\centering
	\includegraphics[width=1\columnwidth]{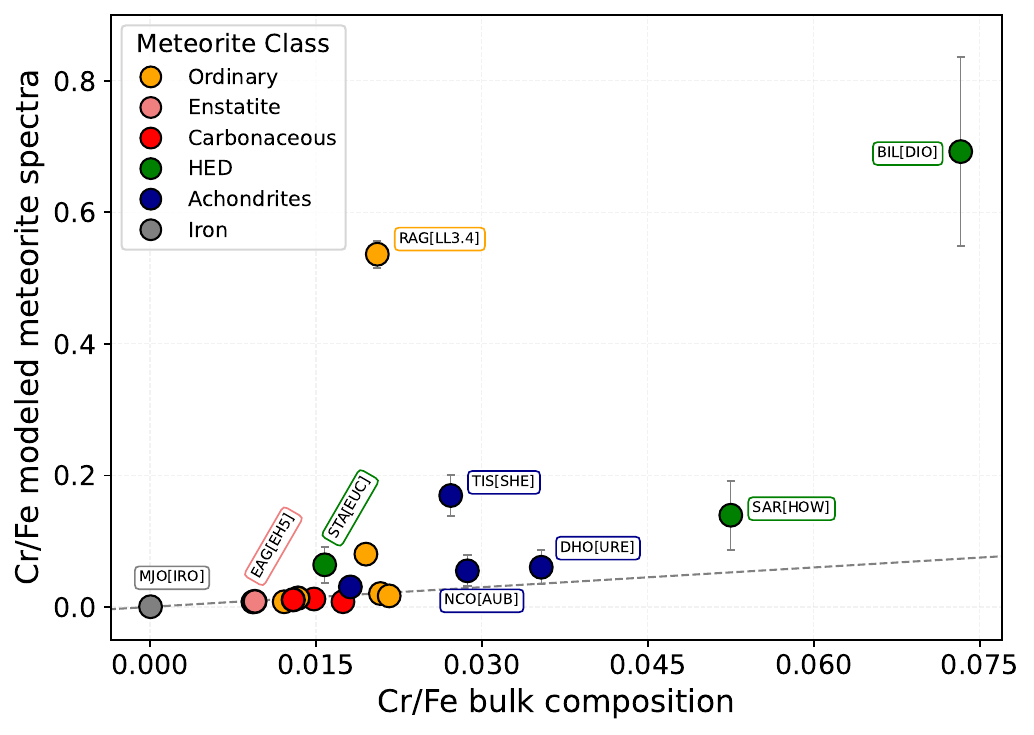}
	\caption{Calculated Cr/Fe mass ratios obtained from spectral modeling of tested meteorites plotted against literature bulk values \citep{2024A&A...689A.323M}. Meteorite classes are distinguished by different colors. The dashed line denotes the 1:1 reference (y = x), corresponding to expected agreement between modeled and bulk Cr/Fe ratios.}
	\label{Cr_to_Fe}
\end{figure}

\begin{figure}
	\centering
	\includegraphics[width=1\columnwidth]{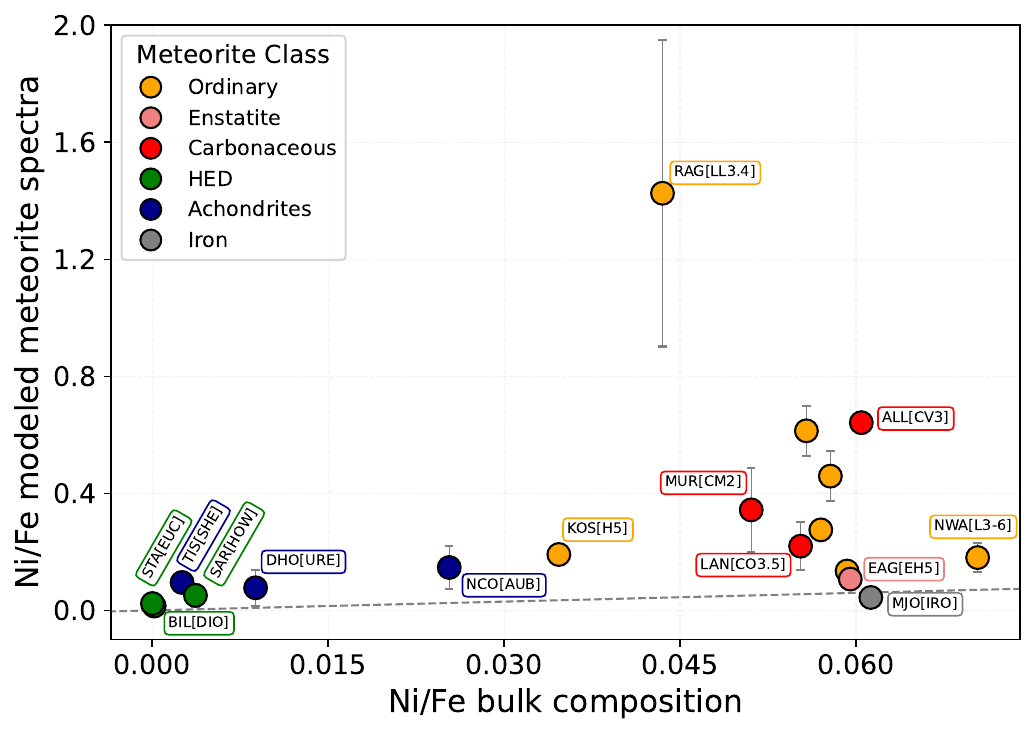}
	\caption{Comparison of calculated Ni/Fe mass ratios derived from spectral modeling versus literature bulk values \citep{2024A&A...689A.323M} for the tested meteorites. Different colors correspond to different meteorite classes. The dashed line denotes the 1:1 reference (y = x), corresponding to expected agreement between modeled and bulk Ni/Fe ratios.}
	\label{Ni_to_Fe}
\end{figure}

\begin{figure}
	\centering
	\includegraphics[width=1\columnwidth]{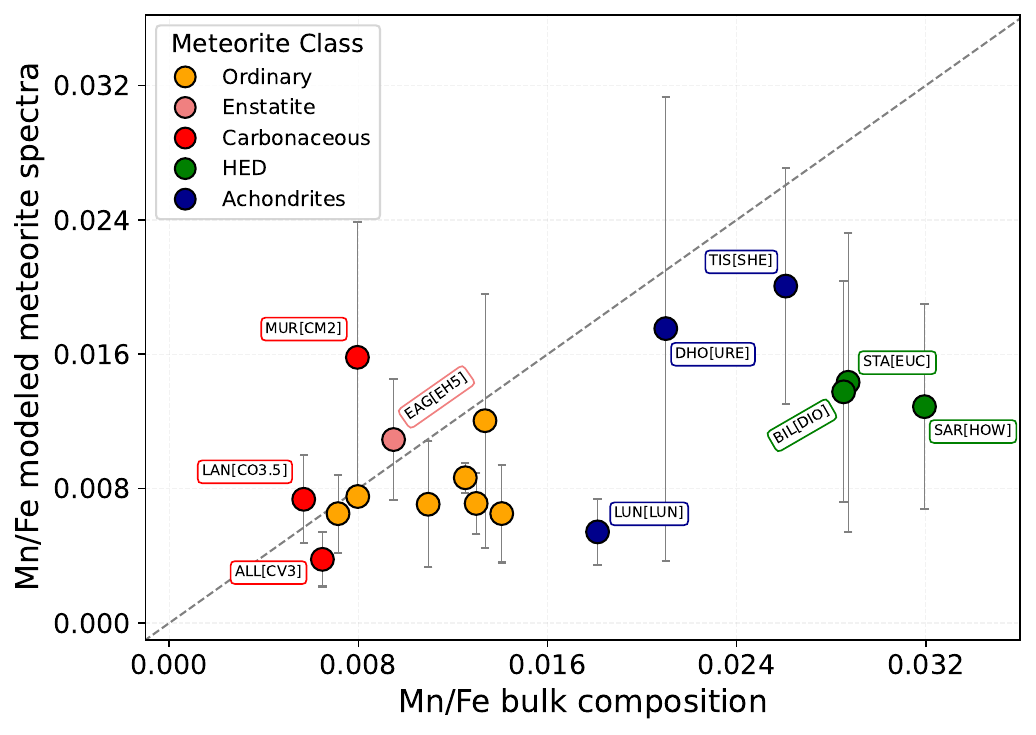}
	\caption{Comparison of spectrally derived Mn/Fe mass ratios from meteorite samples with published bulk composition data \citep{2024A&A...689A.323M}. The color coding represents different meteorite groups. The dashed line denotes the 1:1 reference (y = x), corresponding to expected agreement between modeled and bulk Mn/Fe ratios. The Norton County (aubrite) meteorite is omitted for clarity, as its comparatively large values (x = 0.051, y = 0.091) fall outside the range of the other data points.}
	\label{Mn_to_Fe}
\end{figure}

\begin{figure}
	\centering
	\includegraphics[width=1\columnwidth]{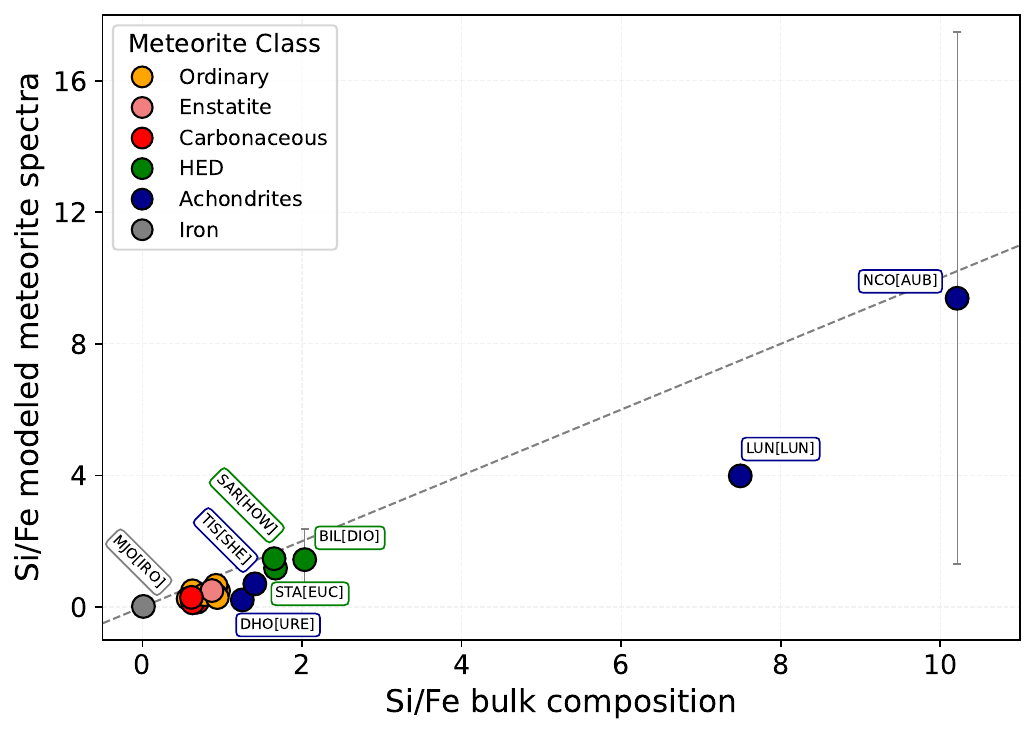}
	\caption{Comparison of calculated Si/Fe mass ratios derived from spectral modeling versus literature bulk values \citep{2024A&A...689A.323M} for the tested meteorites. Different colors correspond to different meteorite classes. The dashed line denotes the 1:1 reference (y = x), corresponding to expected agreement between calculated and bulk Si/Fe ratios.}
	\label{Si_to_Fe}
\end{figure}

\begin{figure}
	\centering
	\includegraphics[width=1\columnwidth]{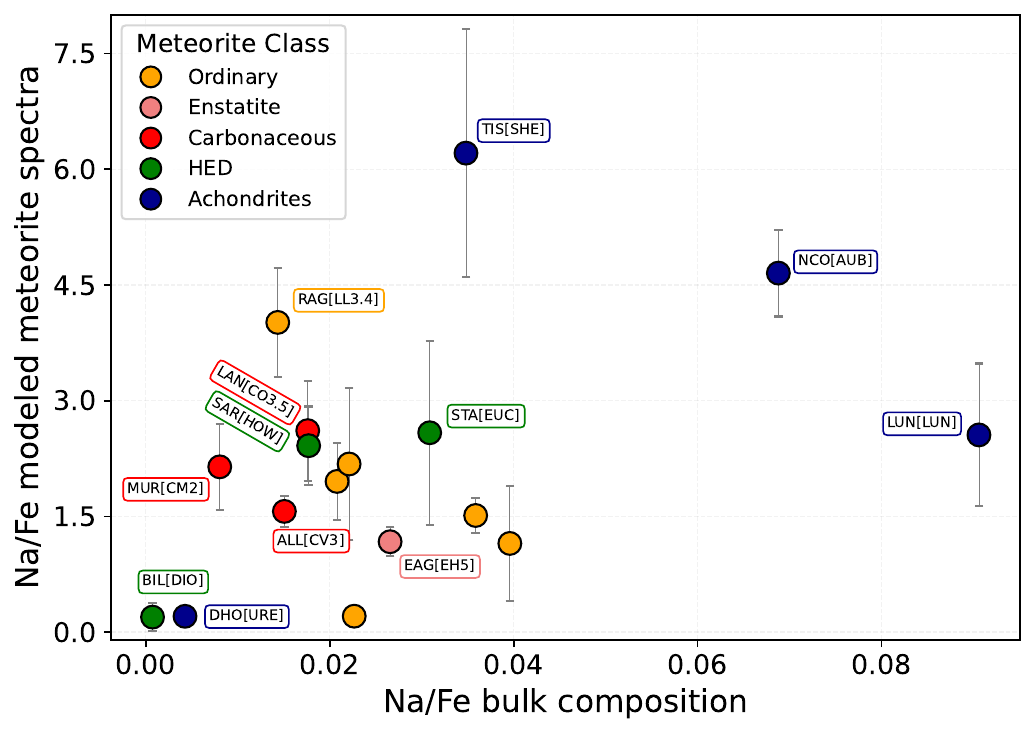}
	\caption{Modeled versus literature Na/Fe \citep{2024A&A...689A.323M} mass ratios for the analyzed meteorite samples. Colors indicate specific meteorite classes.}
	\label{Na_to_Fe}
\end{figure}

\begin{figure}
	\centering
	\includegraphics[width=1\columnwidth]{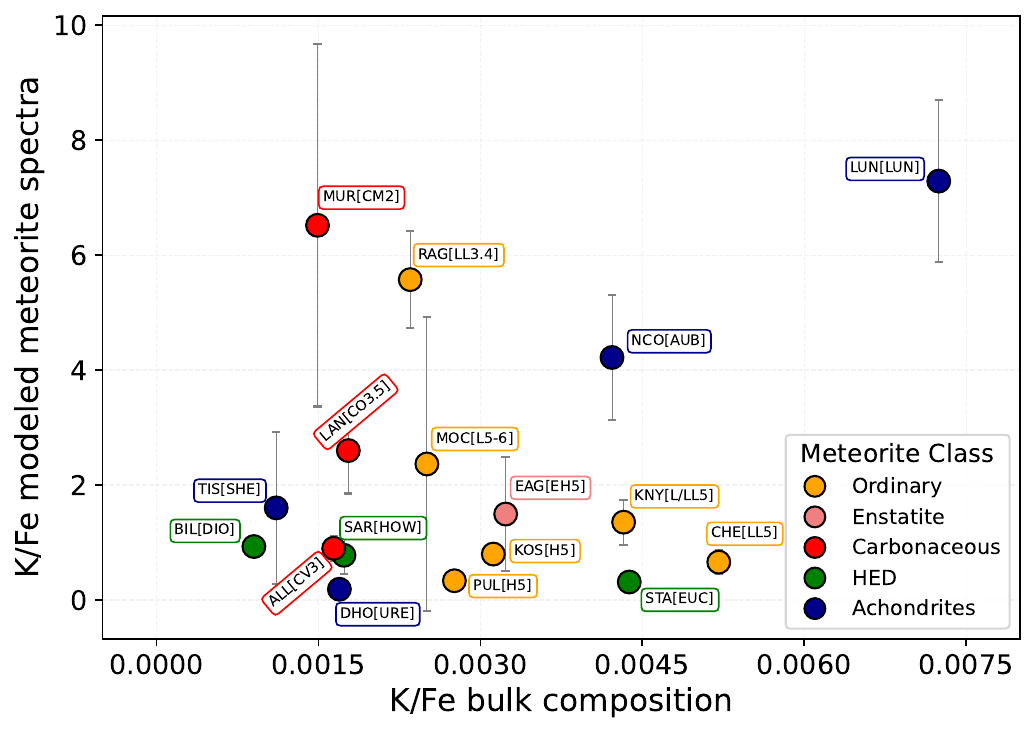}
	\caption{Modeled versus literature K/Fe mass ratios \citep{2024A&A...689A.323M} for the analyzed meteorite samples. Colors indicate specific meteorite classes.}
	\label{K_to_Fe}
\end{figure}

The comparison between calculated elemental mass ratios from the modeled spectra of ablated meteorites and the literature bulk values is illustrated in Figures \ref{Fe_to_Mg} to \ref{K_to_Fe}. For the majority of moderately volatile elements, we observe overall systematic discrepancies, while a general correlation is still present and broadly preserves the relative trends among meteorite classes.

The most prominent discrepancy among these moderately volatile elements is observed in the Fe/Mg mass ratio (Fig. \ref{Fe_to_Mg}). This discrepancy can be illustrated on the Fe/Mg ratios in our dataset of carbonaceous chondrites. Their bulk ratios typically range from 1.5 to 1.8, whereas the ratios derived from the spectra of ablated meteorites are systematically higher, falling in the range of 25.5 to 49.5. This corresponds to an apparent depletion of Mg relative to Fe by an average factor of $\approx$\ 20. Despite this strong depletion of Mg in the radiating vapor, variations in the bulk Fe/Mg ratio among different meteorite groups are still qualitatively reflected in the spectral data. The largest uncertainties in the calculated ratios are found for the carbonaceous chondrite Murchison (CM2) and eucrite Stannern meteorites, primarily because the spectral fits for these samples proved highly sensitive to the selected wavelength range. The physical interpretation and implications of these results are discussed in Section \ref{eqp_vap_model}.

Systematic discrepancies are also evident for Cr and Ni, as shown in Figures \ref{Cr_to_Fe} and \ref{Ni_to_Fe}. In contrast to Mg, both elements exhibit enhancement relative to Fe when compared to their published values. The most prominent outlier in both plots is the ordinary chondrite Ragland (LL3.4), which exhibits an extreme enhancement, with Cr/Fe and Ni/Fe ratios reaching $\approx$\ 0.54 and $\approx$\ 1.43, respectively, compared to published bulk values of $\approx$\ 0.02 and $\approx$\ 0.04. This anomalous value can be attributed to Ragland being a terrestrial find, for which moderate to high weathering and a brecciated internal structure are known to cause compositional irregularities \citep{1986Metic..21..217R}. A distinct deviation is also observed for the diogenite Bilanga, particularly in the Cr/Fe ratio. The Bilanga spectrum shows unusually intense Cr emission lines (mainly Cr I-1 and Cr I-7) that may have reached partially saturation. Consequently, the final spectral fit for this meteorite carries a higher degree of uncertainty, which could have led to a slight overestimation of the derived Cr abundance, as the spectral fitting was highly sensitive to the chosen wavelength interval.

In contrast to the prominent discrepancies observed for Fe/Mg, Cr/Fe, and Ni/Fe, the Mn/Fe and Si/Fe mass ratios calculated from spectral modeling show overall good agreement with published bulk composition data (Figures \ref{Mn_to_Fe} and \ref{Si_to_Fe}). For both elements, the majority of meteorites are distributed close to the 1:1 reference, indicating that the calculated mass ratios are broadly consistent with published values. The most notable deviation is observed for HED (Howardite–Eucrite–Deiogenite) meteorites in the Mn/Fe comparison (Figure \ref{Mn_to_Fe}), where the calculated ratios are on average lower than the corresponding bulk values by a factor of $\approx$\ 2. Aside from this group-specific offset, no significant systematic enhancement or depletion is observed for Mn/Fe or Si/Fe across the analyzed meteorites. It is important to note a methodological limitation regarding the determination of Si abundance. The analysis relied on a single well-observed Si I-3 emission line at 390.55 nm, which lies in a spectral region with lower sensitivity and poorer calibration. As a result, the fit for this line could not be performed with the same precision as for other species. Instead, the Si abundance was derived by manually adjusting the fit to match the intensity of the surrounding Fe I lines, which may introduce a larger uncertainty in the calculated Si/Mg ratios compared to other elements.

Finally, we address the volatile elements Na and K, which exhibit the most significant discrepancies relative to published bulk compositions (Figs. \ref{Na_to_Fe} and \ref{K_to_Fe}). Both elements show strongly enhanced abundances relative to Fe as derived from the spectral modeling. For carbonaceous chondrites, bulk Na/Fe ratios typically range from 0.008 to 0.018. In contrast, our calculated abundance ratios lie between 1.6 and 2.6, corresponding to an average enhancement factor of approximately two orders of magnitude ($\approx$\ 170). An even larger discrepancy is observed for K. While bulk K/Fe ratios are on the order of $\approx$\ 0.0015 for carbonaceous chondrites, the resulting spectral ratios range from 0.9 to 6.5, implying an enhancement of up to three orders of magnitude. It is important to note the methodological challenges associated with fitting the emission lines of these elements, which may contribute to the absence of a clear correlation trend when compared to the moderately volatile elements. The strongest Na I lines (Na I-1 doublet near 589 nm), are highly saturated in most spectra (see the spectrum profile of the Knyahinya meteorite in Fig. \ref{KNY_profiles}). Consequently, the Na abundance was derived using the weaker Na I-6 and Na I-9 multiplets, although the theoretical model often struggled to reproduce the intensities of both multiplets simultaneously. Similarly, for K, the strongest lines of K I-1 at $\approx$\ 766.5 and $\approx$\ 770.0 nm were partially saturated in some cases, and the fit was therefore also constrained using the weaker K I-3 line at $\approx$\ 404.4 nm. These saturation effects, together with the reliance on weaker emission lines, may result in larger uncertainties in the calculated Na and K abundances.

Despite the measurement uncertainties associated with volatile elements, the overall trend of plasma vapor is consistent and physically indicative. The plasma is chemically dominated by volatile elements such as Na and K relative to Fe and is significantly depleted in the moderately volatile Mg, while highly refractory elements such as Al, Ca, and Ti are entirely absent. These findings strongly suggest incomplete vaporization of the ablating material under specific laboratory conditions.

\subsection{Comparison with equilibrium vaporization model}
\label{eqp_vap_model}

Our results are consistent with the equilibrium vaporization model described by \citet{2004EM&P...95..413S}, which predicts that during the initial stages of ablation or under insufficient heating conditions, volatile elements are released preferentially, followed by less volatile elements, leaving the refractory species in the residual melt.

\begin{figure}
	\centering
	\includegraphics[width=1\columnwidth]{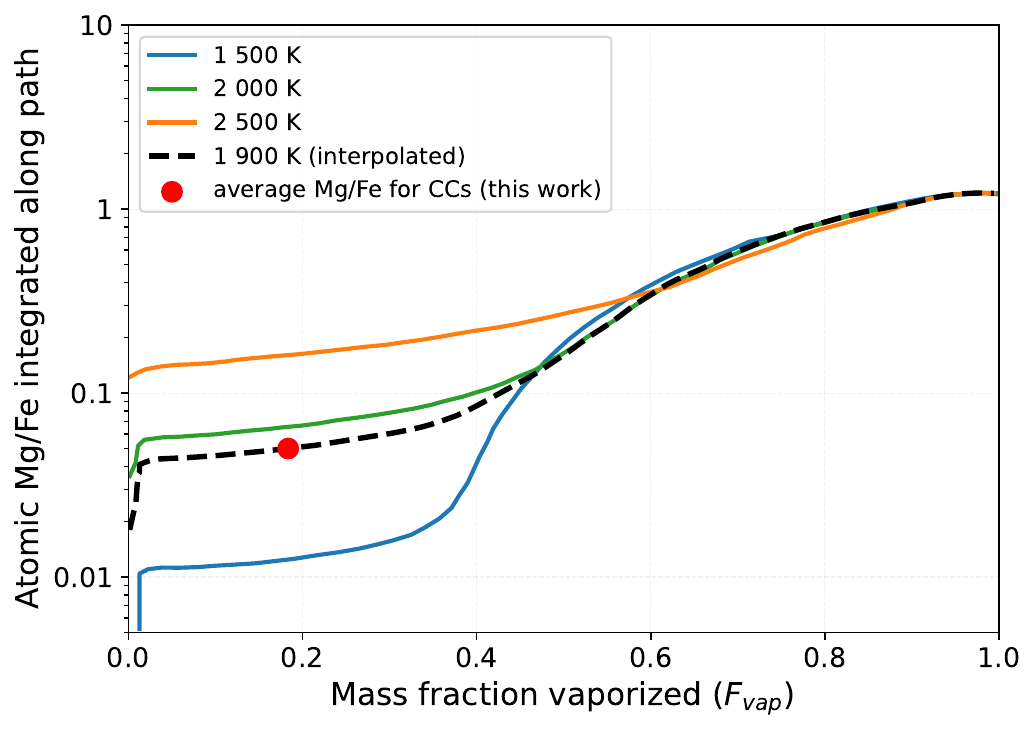}
	\caption{Integrated atomic Mg/Fe ratio as a function of the mass fraction vaporized for a CI chondritic melt at temperatures of 1500 K, 2000 K, and 2500 K, reproduced from Figure 1 (b) in \citet{2004EM&P...95..413S}. The dashed curve shows an interpolated solution for T=1900 K, corresponding to the average surface temperature measured for carbonaceous chondrites in laboratory experiments \citep{2024Icar..40815867L}. The red marker indicates the average Mg/Fe abundance ratio derived from our spectral modeling of carbonaceous chondrites. It corresponds to a vaporized mass fraction of about 20\% according to the equilibrium vaporization model.}
	\label{Vap_model}
\end{figure}

A comparison with MAGMA code simulations for CI chondritic and eucritic materials \citep{2004EM&P...95..413S} supports this interpretation. Their model shows that, at a fixed melt temperature, the atomic Na/Fe ratio in the vapor is significantly enhanced relative to solar values when only a small fraction of the melt has vaporized. This is consistent with the high Na/Fe ratios observed in all our laboratory spectra of ablated meteorites. Furthermore, the model enables us to estimate the degree of vaporization using moderately volatile elements. For example, our spectral modeling of carbonaceous chondrites indicates an average Mg/Fe ratio of $\approx$\ 0.05. For clarity, Fig. \ref{Vap_model} illustrates the Mg/Fe theoretical curves of \citet{2004EM&P...95..413S} for CI chondritic melt, reconstructed and extended with an interpolated solution at T $\approx$\ 1900 K corresponding to the average surface temperature of carbonaceous chondrites measured in laboratory experiments \citep{2024Icar..40815867L}. According to this model, such Mg/Fe depletion corresponds to a scenario in which roughly 20\% of the material has vaporized. In contrast, for Na/Fe, the same model at T $\approx$\ 1900 K indicates that the average value of $\approx$\ 2.1 derived from our spectral modeling of carbonaceous chondrites corresponds to a substantially lower vaporized mass fraction of only $\approx$\ 3\%. Such a high Na/Fe ratio at low vaporization degrees may indicate preferential Na release from subsurface layers into the vapor, a process discussed in more detail below. Although the exact vaporization rate varies by material type, our eucrite sample exhibits qualitatively similar trends which are consistent with the specific curves modeled for eucritic materials. Finally, the model also predicts the absence of emission from refractory elements (Ca, Al) at this temperature, which aligns with our results.

It is important to consider the underlying assumptions and limitations of the MAGMA model in the context of our experiments. According to \citet{2004EM&P...95..413S}, the relative elemental abundances depend on the mass fraction of the vaporized material, the temperature of the melt, and the specific mineralogy. However, the model explicitly assumes that the vaporizing material does not interact with the surrounding atmosphere. Furthermore, the applicability of the equilibrium assumption is size-dependent. While small micrometeoroids (100–1000 $\mu$m) often evaporate completely, cm-sized and larger bodies typically develop only thin surface melt layers. In this regime, heat conduction and diffusion play significant roles. Heat transfer warms the subsurface layers, potentially causing volatile elements to diffuse toward the surface and vaporize preferentially. Consequently, for such larger bodies, chemical equilibrium within the meteoroid itself becomes a less valid assumption. Nevertheless, even under kinetically constrained conditions, the relative order of elemental release follows thermodynamic volatility. Although the vaporization fraction estimated from the model for our data is approximate, the resulting spectral characteristics are consistent with the volatility dependent fractional process introduced by \citet{2004EM&P...95..413S}.

This volatility-driven process is analogous to the ablation observed in meteor spectra. During the early stages of entry at high altitudes, the meteoroid surface experiences relatively moderate heating, leading to the preferential release of volatile species. Consequently, spectra from the upper parts of the meteor trajectory are often dominated by meteoric Na I lines, while emissions from less volatile elements remain weaker \citep{1999M&PS...34..987B}. As the meteoroid penetrates deeper into the atmosphere, the temperature rises significantly, enabling the efficient vaporization of silicates and  partial vaporization of refractory elements. For sufficiently large asteroidal meteoroids capable of deep penetration, the ablation of refractory material continues into the terminal phases of flight, as it was the case of the Bene\v{s}ov bolide \citep{1996Icar..121..484B}. Since our plasma wind tunnel experiments simulated entry conditions corresponding to an altitude of $\sim$80\,km, the observed spectral composition (rich in volatiles but depleted in refractories) is consistent with the incomplete vaporization typical for this high-altitude entry, where the heat transfer to the meteoroid is not yet sufficient to fully vaporize less volatile elements.

\section{Conclusions}
\label{Summary}

We present the first comprehensive validation of radiative-transfer-based compositional analysis using a diverse set of 22 laboratory-ablated meteorites as meteor analogs under controlled conditions. Our analysis resulted in plasma temperatures in the range of 5220–5810 K for individual meteorite samples, slightly higher than those typically reported for the main spectral component of fireballs. Together with the derived Fe I column densities (on the order of $\sim$\num{e14}cm\textsuperscript{-2}) and damping constants ($\sim$\num{e9}s\textsuperscript{-1}), these results indicate that the observed emission is, in most cases, consistent with an optically thick plasma, which requires accounting for self-absorption effects.

Comparison between the derived elemental abundances (true mass ratios) and the known bulk composition of the analyzed meteorites reveals systematic deviations. Relative to Fe, the radiating vapor is significantly enriched in volatile elements such as Na and K, while the moderately volatile element Mg shows depletion. In addition, refractory elements (Al, Ca, and Ti), which commonly appear in meteor spectra at lower altitudes, are absent from the laboratory emission spectra. These results are consistent with the equilibrium vaporization model \citep{2004EM&P...95..413S} and indicate that the energy input during laboratory simulations was insufficient to achieve the full vaporization of the meteoritic material typically observed in meteor spectra, where even refractory species undergo at least partial vaporization.

Our results therefore confirm the importance of incomplete evaporation during the meteoroid ablation. This effect complicates quantitative meteor spectroscopy. Laboratory or theoretical studies of the process of evaporation of meteoritic materials are encouraged to describe
this effect for more elements than it was done by \citet{2004EM&P...95..413S}.

Despite incomplete vaporization, our modeling clearly identifies systematic spectral differences among meteorite classes that reflect their distinct chemical composition. These findings demonstrate that radiative transfer modeling effectively characterizes the radiating plasma, enhances the precision of meteoroid classification when more events are mutually compared, and provides essential insights into volatility-driven processes during atmospheric entry, thereby strengthening our ability to infer meteoroid composition from spectral observations.

\section*{Declaration of competing interest}
The authors declare that they have no known competing financial interests or personal relationships that could have appeared to influence the work reported in this paper.

\section*{Data availability}
The spectral data from laboratory experiments of ablated meteorites are publicly available on the University of Stuttgart data repository DaRUS \citep{DARUS-5100_2025}.

\section*{Acknowledgements}

We thank the High Enthalpy Flow Diagnostics Group team of the Institute of Space Systems, University of Stuttgart for the meteorite experiments. Researcher A. Pisarčíková conducts her research under the Marie Skłodowska-Curie Actions - COFUND project, which is co-funded by the European Union (MERIT - Grant Agreement No. 101081195). This work was supported by ESA under contract no. 4000140\ 012/22/NL/SC/rp and by the Slovak Research and Development
Agency grants VV-MVP-24-0232 and APVV-23-0323.

%

\printcredits

\bibliographystyle{cas-model2-names}

\bibliography{cas-refs}





\end{document}